\documentclass[aps,reprint,prl,letterpaper,floatfix,nobalancelastpage,notitlepage]{revtex4-1}
\usepackage{amsmath}
\usepackage{amsfonts}
\usepackage{graphicx}
\usepackage{bm}
\usepackage{color}
\usepackage{braket}
\usepackage{bbold}
\usepackage[utf8]{inputenc}
\usepackage{hyperref}

\usepackage{color}

\begin{document}

\title{Sampling asymmetric open quantum systems for artificial neural networks}

\author{Oliver Kaestle}
\email{o.kaestle@tu-berlin.de}
\author{Alexander Carmele}
\affiliation{Technische Universit\"at Berlin, Institut f\"ur Theoretische Physik, Nichtlineare Optik und Quantenelektronik, Hardenbergstra{\ss}e 36, 10623 Berlin, Germany}
\date{\today}

\begin{abstract}
While established neural network approaches based on \textit{restricted Boltzmann machine} architectures and Metropolis sampling methods are well suited for symmetric open quantum systems, they result in poor scalability and systematic errors for setups without symmetries of translational invariance, independent of training parameters such as the sample size. To overcome this representational limit, we present a \textit{hybrid sampling} strategy which takes asymmetric properties explicitly into account, achieving fast convergence times and high scalability for asymmetric open systems, underlining the universal applicability of artificial neural networks.
\end{abstract}

\maketitle

Recent advances in the description of quantum states using artificial neural networks~\cite{Carleo2017, DengSarma2017, DengSarmaX2017, Glasser2018, Torlai2018a, Schmitt2020, Burau2020} have opened up new prospects in the field of Markovian open quantum systems. They allow for a direct accessibility of stationary states by application of a variational principle~\cite{Cui2015, Weimer2015}. Combining their nearly limitless potential for parallelization with accurate information compression by Metropolis sampling of possible system configurations in a Markov chain Monte Carlo approach, simulations of very large system sizes become feasible~\cite{Metropolis1953, Robert2004, Kampen2007, SchuldPetruccione2018}.
Specifically, the \textit{restricted Boltzmann machine} (RBM) architecture has emerged as a default network structure for the mapping of the density matrix~\cite{Carleo2017, Torlai2018}. On its basis, efficient simulations of symmetric and periodic open systems have been fathomed, covering calculations for both stationary states~\cite{Yoshioka2019, Vicentini2019} and real-time evolution dynamics~\cite{Hartmann2019, Nagy2019}.

\begin{figure}[b]
\centering
\includegraphics[width=\linewidth]{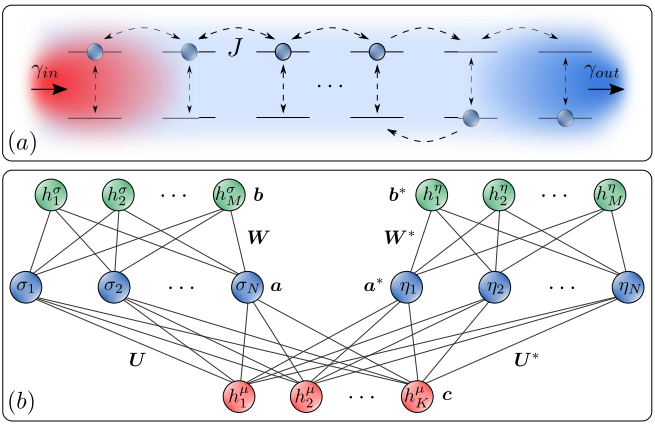}
\caption{(a) Sketch of a boundary-driven isotropic Heisenberg chain with $\gamma^1_{in}>\gamma^1_{out}$, $\gamma^N_{out}>\gamma^N_{in}$ and $\gamma^i_{in}=\gamma^i_{out}$ in the bulk. (b) RBM realization of the \textit{neural density operator}, featuring a visible layer (blue), two auxiliary hidden layers (green) and an ancillary mixing layer (red) connecting the left and right side with training parameters $\bm{\vartheta}=(\bm{a},\bm{b},\bm{c},\bm{W},\bm{U})$.}
\label{fig:system}
\end{figure}

In this Letter, we demonstrate that commonly applied sampling techniques are unsuitable for the training of RBMs representing \textit{asymmetric} open quantum systems, i.e., systems without symmetries of translational invariance.
As an example setup, we calculate the stationary states of a boundary-driven isotropic Heisenberg spin chain with open boundary conditions~\cite{Prosen2011, Znidaric2010, Znidaric2011, Karevski2013, Znidaric2016, Droenner2017, Katzer2020, Finsterhoelzl2020, Finsterhoelzl2020a, Wichterich2007, Popkov2013, Cai2013, Prosen2009, Prosen2015, Xu2018, Mendoza2019, Popkov2020}. First, we provide converging results for a symmetric scenario and further improve computational efficiency by proposing an adjustment to the default Metropolis sampling approach. Secondly, we demonstrate that typical Hilbert space sampling~\cite{Metropolis1953, Hartmann2019, Nagy2019, Vicentini2019} constitutes a systematic overestimation of system asymmetries independent of training parameters such as sample size.
As a solution, we present a \textit{hybrid sampling} design to incorporate asymmetric system properties during compression, combining the accuracy of exact Hilbert space mapping for select sites and typical sampling of the remaining system. Pushing the representational limits of the RBM, we establish a powerful new ansatz for neural network optimization, maintaining fast convergence and high scalability potential at decreased noise.

%%%%%%%%%%%%%%%%%%%%%%

\textit{Model.---} 
To study the representational power of the RBM for asymmetric systems, we consider the model system sketched in Fig.~\ref{fig:system}(a). An isotropic Heisenberg chain featuring $N$ sites, next-neighbor coupling and open boundary conditions is subject to both incoherent dissipation and driving on all sites. The corresponding Hamiltonian is given by~\cite{Prosen2011, Znidaric2010, Znidaric2011, Karevski2013, Znidaric2016, Droenner2017, Katzer2020, Finsterhoelzl2020, Finsterhoelzl2020a, Wichterich2007, Popkov2013, Cai2013, Prosen2009, Prosen2015, Xu2018, Mendoza2019, Popkov2020}
\begin{equation}
H/\hbar =\dfrac{J}{4} \sum_{i=1}^{N-1} \left(\sigma_i^x \sigma_{i+1}^x + \sigma_i^y \sigma_{i+1}^y + \sigma_i^z \sigma_{i+1}^z \right),
\label{eq:H_chain}
\end{equation}
with $\sigma$ denoting Pauli spin matrices and $J$ the next-neighbor coupling amplitude. Together with the Lindblad dissipators
$\mathcal{D} [\sqrt{\gamma^i_{out}/2} \sigma_i^-] \rho = \gamma^i_{out} ( \sigma_i^- \rho \sigma_i^+ - \{ \sigma_i^+ \sigma_i^-, \rho \}/2 )$
and
$\mathcal{D} [\sqrt{\gamma^i_{in}/2} \sigma_i^+] \rho = \gamma^i_{in} ( \sigma_i^+ \rho \sigma_i^- -  \{ \sigma_i^- \sigma_i^+, \rho \} /2 )$,
imposing incoherent excitation and decay on sites $i$ at rates $\gamma^i_{in}$ and $\gamma^i_{out}$, the unfolding time evolution dynamics in Liouville space is prescribed by $\dot{\rho} = \mathcal{L} \rho = -i[H/\hbar,\rho] + \sum_{i=1}^{N} \{ \mathcal{D} [\sqrt{\gamma^i_{out}/2} \sigma_i^-] + \mathcal{D} [\sqrt{\gamma^i_{in}/2} \sigma_i^+] \} \rho$.
Using a brute force fourth order Runge Kutta approach, system lengths of typically $N \lesssim 14$ are accessible without requiring excessive computational resources and in the following serve as a benchmark to study the representational limits of the RBM.

%%%%%%%%%%%%%%%%%%%%%%

\textit{Neural network implementation.}---
The core idea behind neural network descriptions for open quantum systems is to simulate the density matrix by a probabilistic architecture, replacing $2^{2N}$ elements $\bra{\bm{\sigma}} \rho \ket{\bm{\eta}}= \bra{\sigma_1, \ldots, \sigma_N} \rho \ket{\eta_1,\ldots,\eta_N}$ for a system of $N$ sites by a set of variational parameters to be optimized.
In recent breakthroughs, the RBM has been established as a default neural network architecture for the approximate mapping of the density matrix of symmetric open quantum systems~\cite{Carleo2017, Torlai2018, Hartmann2019, Nagy2019, Vicentini2019, Yoshioka2019, Carleo2019, Vieijra2020, Carleo2018, Cheng2018}. It constitutes a model distribution called \textit{neural density operator} (NDO), which consists of a visible layer of $2N$ sites $\bm{\sigma}$ and $\bm{\eta}$ representing the configuration of the left and right side of the density matrix, two auxiliary hidden layers with $M$ sites $\bm{h}^\sigma$ and $\bm{h}^\eta$ each, and an ancillary mixing layer of $K$ neurons $\bm{h}^\mu$ [see Fig.~\ref{fig:system}(b)]. Tracing out the hidden and ancillary degrees of freedom, the elements of the NDO read
\begin{align}
&\rho_{\bm{\vartheta}} (\bm{\sigma}, \bm{\eta}) = 8 \exp \left( \sum_{i=1}^N a_i \sigma_i \right)  \exp \left( \sum_{i=1}^N a_i^* \eta_i \right)  \nonumber \\
& \times \prod_{m=1}^M \cosh \left( b_m + \sum_{i=1}^N W_{mi} \sigma_i \right) \cosh \left( b_m^* + \sum_{i=1}^N W_{mi}^* \eta_i \right) \nonumber \\
& \times \prod_{k=1}^K \cosh \left( c_k + c_k^* + \sum_{i=1}^N U_{ki} \sigma_i  + \sum_{i=1}^N U_{ki}^* \eta_i \right),
\label{eq:rbm_ndo_traced}
\end{align}
with a set of complex training parameters $\bm{\vartheta}=(\bm{a},\bm{b},\bm{c},\bm{W},\bm{U})$ split up into real and imaginary parts, resulting in $n_p = 2N + 2M + K + 2MN + 2KN$ elements. The parameters consist of biases $\bm{a}$ for visible units, $\bm{b}$ for auxiliary hidden units and $\bm{c}$ for the ancillary mixing layer, respectively. The complex weights $\bm{W}$ connect visible units of $\bm{\sigma}$ and $\bm{\eta}$ to their auxiliary hidden counterparts $\bm{h}^\sigma$ and $\bm{h}^\eta$. Weights $\bm{U}$ couple visible units to the ancillary mixing layer $\bm{h}^\mu$ [see Fig.~\ref{fig:system}(b)].
Our goal is to approximate the unknown density matrix $\rho$ by the NDO $\rho_{\bm{\vartheta}}$ via iterative variational optimization of the parameter set $\bm{\vartheta}$. During each learning iteration, input training data is produced by drawing $N_s$ samples of possible left and right system spin configurations.
Following the stochastic reconfiguration approach~\cite{Sorella1997, Sorella2007, BeccaSorella2017}, both the partition function and system observables can be approximated as statistical expectation values over all samples. For the normalized occurrence probability of the $n$-th drawn sample $(\bm{\sigma}_n,\bm{\eta}_n)$, we approximate the NDO partition function by summing over all $N_s$ samples drawn during one iteration, yielding the probability $\tilde{p}_{\bm{\vartheta}}(\bm{\sigma}_n,\bm{\eta}_n) = | \rho_{\bm{\vartheta}} (\bm{\sigma}_n,\bm{\eta}_n) |^2 / [\sum_{n=1}^{N_s} | \rho_{\bm{\vartheta}} (\bm{\sigma}_n,\bm{\eta}_n) |^2 ]$. %
Expectation values of diagonal observables are then estimated as statistical averages $\braket{X(\bm{\sigma},\bm{\sigma})} \approx \langle \langle X(\bm{\sigma},\bm{\sigma}) \rangle \rangle_q = \sum_{n=1}^{N_s} \tilde{q}_{\bm{\vartheta}}(\bm{\sigma}_n) \sum_{\bm{\xi}} X(\bm{\sigma}_n,\bm{\xi}) \rho_{\bm{\vartheta}}(\bm{\xi}, \bm{\sigma}_n) /  \rho_{\bm{\vartheta}} (\bm{\sigma}_n,\bm{\sigma}_n)$~\cite{Sorella1997, Sorella2007, BeccaSorella2017, Hartmann2019, Vicentini2019, Nagy2019}, with probabilities of diagonal samples $\tilde{q}_{\bm{\vartheta}}(\bm{\sigma}_n) = \rho_{\bm{\vartheta}} (\bm{\sigma}_n, \bm{\sigma}_n) / [\sum_{n=1}^{N_s} \rho_{\bm{\vartheta}} (\bm{\sigma}_n, \bm{\sigma}_n) ]$~\footnote{The diagonal probability distribution $\tilde{q}_{\bm{\vartheta}}(\bm{\sigma})$ is introduced for improved efficiency, since diagonal observables are figures of merit. They are calculated by additionally drawing $N_s$ diagonal samples $(\bm{\sigma}_n,\bm{\sigma}_n)$ during each iteration. Hence, $\tilde{q}_{\bm{\vartheta}}(\bm{\sigma})$ is used for the calculation of diagonal observables and $\tilde{p}_{\bm{\vartheta}}(\bm{\sigma},\bm{\eta})$ for the training of the network.}.

Our training objective is to find the stationary state of the considered model system, prescribed by the condition $\dot{\rho} = \mathcal{L} \rho = 0$.
To this end, we define a cost function $C(\bm{\vartheta}) = \left\| \mathcal{L} \rho_{\bm{\vartheta}} \right\|_2^2$~\cite{Vicentini2019, Nagy2019}. Using the $N_s$ samples as input, the neural network is trained via the standard stochastic gradient descent procedure, updating the variational parameters $\bm{\vartheta}$ during each iteration $t \rightarrow t+1$ via $\vartheta_l^{(t+1)} = \vartheta_l^{(t)} - \nu \nabla_{\vartheta_l} C(\bm{\vartheta}^{(t)})$ at learning rate $\nu$~\cite{SchuldPetruccione2018}. The variational parameters are initialized at small nonzero random values, $\vartheta_l^{(0)} \in [-0.01,0.01] \backslash \{ 0\}$.
The corresponding cost function gradient is evaluated as~\cite{Vicentini2019}
\begin{align}
&\nabla_{\vartheta_l} C(\bm{\vartheta}) = 2 \mathrm{Re} \Bigg\{ \sum_{n=1}^{N_s} \tilde{p}_{\bm{\vartheta}} (\bm{\sigma}_n,\bm{\eta}_n) \tilde{\mathcal{\bm{L}}}^\dagger (\bm{\sigma}_n,\bm{\eta}_n) \nonumber \\
&\times \sum_{m=1}^{N_s} \mathcal{\bm{L}} (\bm{\sigma}_n,\bm{\eta}_n, \bm{\sigma}_m, \bm{\eta}_m) \dfrac{\bm{\rho}_{\bm{\vartheta}}(\bm{\sigma}_m, \bm{\eta}_m)}{\bm{\rho}_{\bm{\vartheta}}(\bm{\sigma}_n, \bm{\eta}_n)} O_{\vartheta_l}(\bm{\sigma}_m, \bm{\eta}_m) \nonumber \\
&- \left[ \sum_{n=1}^{N_s} \tilde{p}_{\bm{\vartheta}} (\bm{\sigma}_n,\bm{\eta}_n) O_{\vartheta_l}(\bm{\sigma}_n, \bm{\eta}_n) \right] \nonumber \\
& \times \left[ \sum_{n=1}^{N_s} \tilde{p}_{\bm{\vartheta}} (\bm{\sigma}_n,\bm{\eta}_n) \tilde{\mathcal{\bm{L}}}^\dagger (\bm{\sigma}_n,\bm{\eta}_n) \tilde{\mathcal{\bm{L}}} (\bm{\sigma}_n,\bm{\eta}_n) \right]  \Bigg\},
\end{align}
where we introduced logarithmic derivatives in the form of diagonal matrices with elements $[\bm{O}_{\vartheta_l}]_{\bm{\sigma}_n \bm{\eta}_n, \bm{\sigma}_n \bm{\eta}_n} = O_{\vartheta_l} (\bm{\sigma}_n, \bm{\eta}_n) = \partial [ \ln \rho_{\bm{\vartheta}}(\bm{\sigma}_n,\bm{\eta}_n) ]/ \partial \vartheta_l$, corresponding to the gradients of the NDO with respect to all $l$ elements of the set of variational parameters $\bm{\vartheta}$ and for a specific sample configuration $(\bm{\sigma}_n,\bm{\eta}_n)$.
Moreover, we introduced the estimator of the Liouvillian, $\tilde{\mathcal{L}}( \bm{\sigma}_n, \bm{\eta}_n) := \sum_{\bm{\sigma}_m,\bm{\eta}_m} \mathcal{L} (\bm{\sigma}_n, \bm{\eta}_n, \bm{\sigma}_m, \bm{\eta}_m ) \rho_{\bm{\vartheta}} (\bm{\sigma}_m,\bm{\eta}_m) / \rho_{\bm{\vartheta}} (\bm{\sigma}_n,\bm{\eta}_n)$.

%%%%%%%%%%%%%%%%%%%%%%

\begin{figure}[t]
\centering
\includegraphics[width=\linewidth]{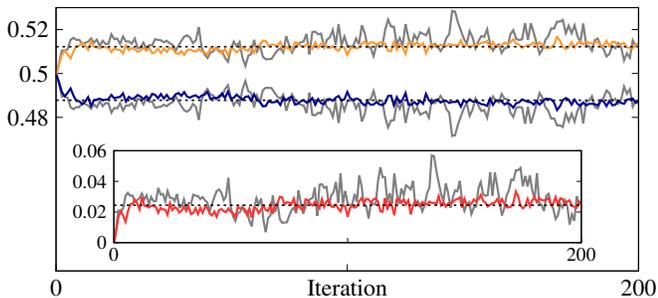}
\caption{RBM estimation for a \textit{symmetric} chain with $N=10$ obtained via regular sampling (grey lines) and the adjusted sampling method showing (colored lines), showing the stationary ground (blue line) and excited state occupations (orange line). In both sampling cases, ground and excited state populations are averaged over all sites. Dashed black lines depict benchmark results. Inset: Steady state magnetization $m_z$.}
\label{fig:adhc}
\end{figure}

\textit{Symmetric systems: Improving sampling efficiency.}---
To demonstrate the power of the neural network approach for symmetric systems, we start by considering a symmetric, isotropic Heisenberg chain with $\gamma^i_{in}=1.05\gamma^i_{out}$ on \textit{all} sites $i$ in case of incoherent driving and open boundary conditions. Similar symmetric setups featuring coherent driving, symmetric dissipation and periodic boundary conditions have been recently realized with RBMs, enabling simulations of large systems with $N \geq 15$~\cite{Vicentini2019, Nagy2019, Hartmann2019}.
Due to exponential growth of the Hilbert space dimension, at these lengths an exact mapping of the density matrix becomes computationally very expensive. The established solution for an efficient compression of configuration space is to employ the Metropolis algorithm~\cite{Metropolis1953}, where a sequence of $N_s$ random samples is drawn based on a Markov chain Monte Carlo method, corresponding to a random walk in Hilbert space~\cite{Robert2004, Kampen2007, SchuldPetruccione2018}. Based on a set of selection rules, a new system configuration is drawn based on the preceding sample and either accepted or rejected with a certain acceptance probability. Here we choose as selection rule to flip the spin of each site at $50\%$ probability.
The acceptance probability is chosen as
\begin{equation}
A(n+1,n) = \mathrm{min} \left[ 1, \dfrac{\tilde{p}_{\bm{\vartheta}}(\bm{\sigma}_{n+1},\bm{\eta}_{n+1})}{\tilde{p}_{\bm{\vartheta}}(\bm{\sigma}_n,\bm{\eta}_n)} \right],
\label{eq:accept}
\end{equation}
with $(\bm{\sigma}_n,\bm{\eta}_n)$ denoting the current sample and $(\bm{\sigma}_{n+1},\bm{\eta}_{n+1})$ the proposed sample. If the proposed sample is rejected, the standard Metropolis algorithm continues with the current sample, resulting in convergent results for the case of symmetric systems and further improving for increasing sample sizes $N_s$.
As a first benchmark, we compare results obtained from the default Metropolis sampling method with benchmark values from a Runge Kutta master equation solution at length $N=10$ and parameters $\gamma_{in}=0.21$, $\gamma_{out}=0.20$ and $J=2\gamma_{in}$. For the neural network, we employ hidden layer densities $M/N=K/N=1$ and $N_s=20000$ samples at a learning rate $\nu=0.1$.
Compared with the required $\approx 500000$ density matrix elements calculated in a brute force implementation, the neural network compresses the information onto $n_p=450$ variational coefficients to be optimized by training.
Grey lines in Fig.~\ref{fig:adhc} show RBM calculations of the mean stationary ground and excited state populations for the symmetric chain and averaged over all $N=10$ sites for improved visibility. The inset shows the steady state magnetization $m_z = (1/N) \sum_{i=1}^N \braket{\sigma_i^z}$. Runge Kutta benchmark values are indicated by dashed black lines, showing overall agreement with the results obtained from regular Metropolis sampling. To achieve improved convergence with a lower standard deviation, the sample size $N_s$ must be increased.

For the following calculations of asymmetric setups, we propose instead an adjusted sampling procedure which breaks the detailed balance condition in the Markov chain and introduces a net stochastic flow in configuration space, resulting in enhanced convergence by suppressing the random walk behavior~\cite{Suwa2010}.
Rather than continuing with the current sample in the event of a rejection, we employ an \textit{accept-only} agenda: To obtain the next current sample, new configurations $(\bm{\sigma}_{n+1},\bm{\eta}_{n+1})$ are proposed until one of them is accepted by the algorithm. As a result, no sample configuration occurs more than once in a row unless it is chosen by the selection rule, considerably reducing autocorrelations in the Markov chain~\footnote{Based on the employed selection rule for the drawing of a new sample, a sample configuration may occur twice in a row at a probability $1/2^N$.}.
Colored lines in Fig.~\ref{fig:adhc} show corresponding symmetric calculations using the adjusted sampling method for the stationary ground and excited states averaged over all sites (blue and orange lines) and the magnetization (red line in inset). While both sampling approaches result in fast convergence, the adjusted \textit{accept-only} strategy leads to overall smoother lineshapes with fewer kinks, even more so for increased chain lengths (not shown): After $100$ iterations, the maximum relative deviation of the regular sampling case from the benchmark is $3.35\%$ against $0.99\%$ for the \textit{accept-only} sampling, corresponding to an improvement ratio $3.35/0.99 \approx 3.38$.

%%%%%%%%%%%%%%%%%%%%%%

\begin{figure}[t]
\centering
\includegraphics[width=\linewidth]{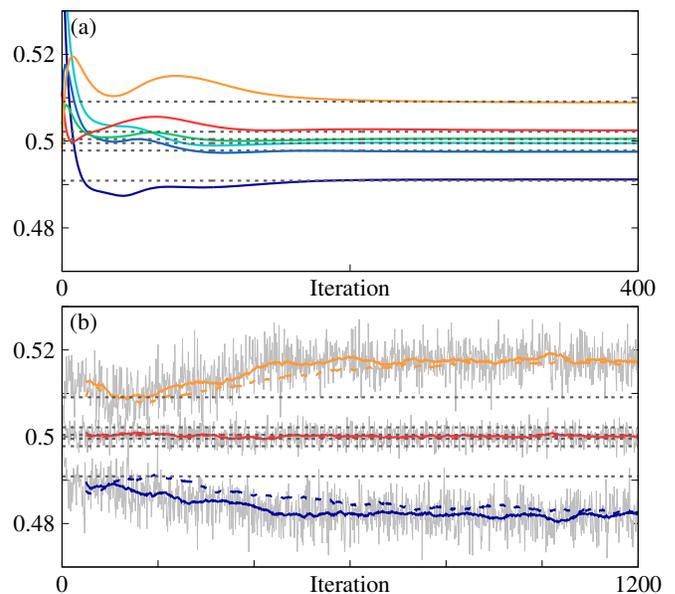}
\caption{(a) Stationary ground state occupations for an \textit{asymmetric} chain with $N=6$ obtained via exact Hilbert space mapping. Dashed grey lines indicate benchmark populations. Dark blue and orange lines depict the left and right boundary ground site occupations. (b) Results obtained from adjusted sampling. Grey lines show raw results, resulting in colored lines when averaged over. Populations of all bulk sites are additionally averaged over (red line). Dashed colored lines show comparison calculations at half the learning rate and twice the hidden layer densities and number of samples.}
\label{fig:bbdhc_reg}
\end{figure}

\textit{Asymmetric systems: Systematic sampling errors.}---
As a next step, we consider the asymmetric case by setting $\gamma^i_{in}=\gamma^i_{out}=0.20$ on all bulk sites $i$, $\gamma^1_{in}=0.21$, $\gamma^1_{out}=0.20$ on the left edge [red shape in Fig.~\ref{fig:system}(a)] and $\gamma^N_{in}=0.20$, $\gamma^N_{out}=0.21$ on the right edge (dark blue shape), constituting a boundary-driven asymmetric chain. We leave all remaining parameters unchanged with respect to the previous symmetric case.
To test the ability of the RBM to model such a system, we start with $N=6$ sites where an exact mapping is unproblematic: Rather than sampling only parts of the corresponding Hilbert space, in Fig.~\ref{fig:bbdhc_reg}(a) \textit{all} possible system configurations are taken into account for each training iteration. As a result, the neural network perfectly recreates the stationary ground state occupations of the asymmetric chain [solid colored lines in Fig.~\ref{fig:bbdhc_reg}(a)], with dashed grey lines indicating benchmark occupations. Hence, the RBM architecture is by itself not limited to symmetric systems, proving its viability for asymmetric systems in case of an exact mapping.

However, the picture drastically changes when regular sampling is applied [see Fig.~\ref{fig:bbdhc_reg}(b)]. For improved visibility, the resulting ground state occupations (grey lines) are averaged over 50 iterations (solid colored lines). Furthermore, for improved visibility all bulk site occupations are averaged over (solid red line). Dashed grey lines again indicate Runge Kutta benchmark occupations.
While all site occupations of the symmetric bulk are in very good agreement with their benchmarks, the steady state populations of the boundary sites overshoot considerably [solid blue and orange lines in Fig.~\ref{fig:bbdhc_reg}(b)]. Crucially, we have ensured that this behavior occurs independently from the choice of hidden layer densities $M/N$ and $K/N$, sample size $N_s$, learning rate $\nu$ and system size $N$. Dashed colored lines show comparison calculations for $\nu=0.05$, $N_s=40000$ and $M/N=K/N=2$.
Moreover, this error occurs for both the regular Metropolis and the adjusted \textit{accept-only} sampling approach and is robust against the specific choice for the selection rules and the acceptance function: Choosing e.g. an exponential acceptance probability $A(n+1,n)=\exp [-\tilde{p}_{\bm{\vartheta}}(\bm{\sigma}_{n},\bm{\eta}_{n})/\tilde{p}_{\bm{\vartheta}}(\bm{\sigma}_{n+1},\bm{\eta}_{n+1})]$ still results in the same behavior.
In direct comparison with Fig.~\ref{fig:bbdhc_reg}(a), this only leaves the sampling method itself as source of error: Caused by the asymmetric nature of the setup, Hilbert space elements with a low probability of occurrence may still have a large impact on the stationary state under certain circumstances. Applying established sampling techniques for symmetric systems in such scenarios results in a systematic error, since spin configurations with higher occurrence probabilities are favored per se [see Eq.~\eqref{eq:accept}].
Hence, we find that common Metropolis-based sampling techniques are unsuitable for the description of such asymmetric open systems. As a solution, in the following we propose a \textit{hybrid sampling} strategy which explicitly takes the asymmetric properties of the considered system into account, enabling efficient and accurate neural network representations of asymmetric open systems.

%%%%%%%%%%%%%%%%%%%%%%

\textit{Hybrid sampling for asymmetric systems.}---
Having confirmed a systematic error during the sampling of asymmetric systems, we present a \textit{hybrid sampling} scheme which combines the accuracy of exact Hilbert space mapping [see Fig.~\ref{fig:bbdhc_reg}(a)] with sampling to ensure computational accessibility [see Fig.~\ref{fig:bbdhc_reg}(b)]. In the considered boundary-driven scenario, the asymmetry is located at the edge sites of the Heisenberg chain.
The core idea behind the hybrid sampling strategy is to manually choose the edge configurations in all samples and leaving only the bulk subjected to the sampling algorithm, as described in detail in the following.
As in the regular sampling case, first a new proposed sample $(\bm{\sigma}_{n+1},\bm{\eta}_{n+1})$ is drawn. To avoid an arbitrary acceptance mechanism, we determine the occurrence probability of the bulk configuration (i.e., the symmetric subsystem) independently from the boundaries by calculating the mean probability over all 16 possible edge configurations for the bulk.
Next, the boundary configuration of the new sample is manually set to one of 16 possible settings. After processing 16 accepted samples, each with a different bulk configuration, a cycle of all possible edge configurations has been completed~\footnote{For the case of diagonal samples $(\bm{\sigma}_n, \bm{\sigma}_n)$, only four different boundary settings are possible}. The procedure is repeated for all $N_s$ samples drawn during each iteration.

\begin{figure}[t]
\centering
\includegraphics[width=\linewidth]{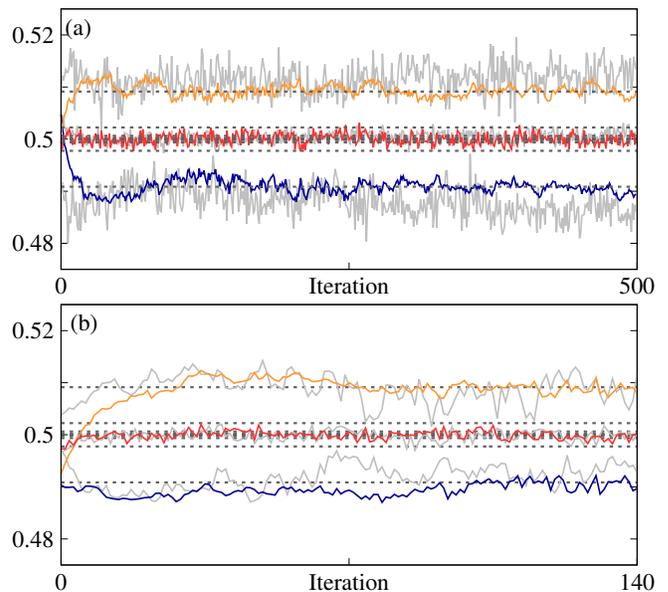}
\caption{Stationary ground state occupations of the boundary-driven \textit{asymmetric} chain, obtained via \textit{hybrid sampling}. (a) Calculations for a chain of $N=10$ sites. Blue and orange lines depict left and right boundary site occupations. Ground state occupations of all bulk sites are averaged over for improved visibility (red line). Dashed grey lines indicate Runge Kutta benchmark populations. Grey lines show comparison calculations with regular sampling. (b) Hybrid sampling calculations for a chain of $N=16$ sites at $N_s=40000$, $\nu=0.05$ (grey lines) and $\nu=0.025$ (colored lines).}
\label{fig:bbdhc_mixed}
\end{figure}

This hybrid strategy of accurate and approximate mapping ensures a thorough sampling of asymmetric system properties, counteracting the systematic overestimation of their impact during regular Hilbert space sampling and compression~\footnote{The mapping of the boundary sites is still approximate. For a truly exact mapping, one would need to remain with the same bulk configuration for all 16 edge settings. However, this comes at the cost of severely decreased bulk convergence at a fixed sample size.}.
The resulting convergence accuracy is only limited by the choice of training parameters. Fig.~\ref{fig:bbdhc_mixed}(a) shows the stationary ground state occupations of the asymmetric Heisenberg chain for $N=10$ sites, $N_s=40000$ and $\nu=0.05$ and otherwise unchanged parameters with respect to Fig.~\ref{fig:bbdhc_reg}(b). Again, all bulk site occupations are averaged over (red line) and once more match their benchmarks (dashed grey lines).
Strikingly, in difference to the regular sampling case (solid grey lines), the stationary ground state occupations of the left and right boundary sites exhibit distinctly reduced fluctuations and no longer overshoot (blue and orange lines). As a result of hybrid sampling, they quickly converge to their benchmark values (dashed grey lines) with a maximum relative deviation of $0.45\%$ after $300$ iterations.
Lastly, Fig.~\ref{fig:bbdhc_mixed}(b) shows calculations for $N=16$ sites, $N_s=40000$, $\nu=0.05$ (grey lines) and $\nu=0.025$ (colored lines), demonstrating the high scalability and performance of the hybrid sampling strategy. As a guide for the eye, we included the Runge Kutta benchmark occupations for $N=10$ (dashed grey lines). As expected, the edge occupations for the longer chain start to strive towards the bulk occupation values, since by increasing the chain length the impact of the asymmetry due to different edge driving decreases.

%%%%%%%%%%%%%%%%%%%%%%

\textit{Conclusion.}---
We have demonstrated that established neural network sampling techniques are unsuitable for the representation of asymmetric open quantum systems due to a systematic overestimation of asymmetric properties during sample selection.
Introducing a hybrid sampling algorithm which combines the accuracy of exact Hilbert space mapping for select sites and efficient compression by sampling of the remaining symmetric subsystem, convergence is achieved with few iterations and decreased noise. Moreover, we have proposed an adjustment to the regular Metropolis algorithm, resulting in improved convergence~\footnote{The numerical realization of the presented sampling strategies is publicly available under \url{https://github.com/okaestle/rbm_sampling}.}.
The presented methods create a novel access point for the tailoring and optimization of artificial neural networks via adaptive sampling strategies while maintaining their high accuracy and performance, making the RBM architecture an ever more powerful and versatile tool for the simulation of open quantum systems.

%%%%%%%%%%%%%%%%%%%%%%

\begin{acknowledgments}
\textit{Acknowledgments.}--- The authors acknowledge support from the Deutsche Forschungsgemeinschaft (DFG) through SFB 910 project B1 (Project No. 163436311).
\end{acknowledgments}

%%%%%%%%%%%%%%%%%%%%%%

%\bibliography{ann_paper}

%

\end{document}